Argyn Kuketayev

Email: jawabean@gwu.edu

First Year "Summer Paper", PhD Program

# The convergence of regional house prices in the USA in the context of the stress testing of financial institutions

**Abstract**

I studied the convergence of regional house prices to national prices in USA by analyzing time-series of house price indices of 9 Census Divisions. I found the evidence of the convergence in some parts of the country using asymmetric unit root tests. The fact that the evidence of the convergence is not present in large parts of the country raises an issue of execution and interpretation of results of Federal Reserve Bank's annual stress testing of the US banking system.

# Contents

# Acknowledgements

Author is grateful to Professor Min Hwang for fruitful discussions and advice, especially the suggestion to use ARMA model for comparisons.

# Introduction

Our study of the convergence of regional house prices in the USA is motivated by two reasons. First, until recently the sentiment among the researchers and practitioners alike was that US housing markets are essentially local and reflect the very specific local market conditions, or to put it bluntly that there is no national housing market in the USA. However, it is clear now that US experienced the nationwide house price bubble that burst in 2007 with dramatic consequences to US economy. This brings up the question whether the notion of national housing market should be considered and there is a long-run convergence between local housing markets; or whether as Greenspan said (see Krishna, 2007) this bubble was really "a collection of bubbles". The existence (or non-existence) of the national housing market in the USA has policy implications because it may impact our understanding of how nation-level economic decisions propagate into regional housing markets.

Our second motivation is of more immediate concern. In 2011 Federal Reserve Bank (FRB) launched the first Comprehensive Capital Analysis and Review (CCAR) (see FRB CCAR and FRB (2012)). As part of this exercise national top-tier banks have to forecast their balance sheets under the macroeconomic stress scenarios that are provided by FRB. For instance, in CCAR 2013 there are 26 variables including the national house price index (HPI). However, without the notion of the national housing market one could argue that national HPI is useless for determination of regional HPIs, which presents a problem to the banks involved in the stress testing, because they usually use local house price indices to forecast mortgage cash flows. Consider Wells Fargo Bank NA that – according to FDIC Call report as of 12/31/2012 – has 452 billion dollars of loans secured by real estate, including over 300 billion dollars secured by 1-4 family residential properties. For a bank like Wells Fargo HPI is a very important variable because of the sheer size of its mortgage portfolio. Hence, understanding the relationship between national and regional HPIs is of critical importance in the context of the stress testing exercise conducted by FRB.

## Literature Review

Until very recently the literature on the convergence of regional US house prices was sparse at best as pointed out by Cameron (2007), perhaps because the housing markets in USA were considered to be local as I mentioned earlier.

There are several papers on price determinants of regional housing markets, which do not consider the nationwide trends. Hwang and Quigley (2006) study the price determinants in 74 metropolitan areas. Capozza et al. (2004) suggest mean-reverting HPI model for metropolitan areas, which is a base for Moody's Economy.com service's regional HPI forecasts, see Chen et al. (2011). Pollakowsky and Ray (1997) examine diffusion of shocks between Census Divisions and between metropolitan statistical areas (MSA). They find that indeed the shocks propagate to neighboring MSAs, and they call this phenomenon *contiguous spatial diffusion*.

Recently, several papers came out that explicitly study the long-run convergence of regional house prices. Fu (2007) decomposes regional HPI moves into national, regional and idiosyncratic factors using OFHEO (FHFA) HPI series on MSA level. The author states that "strictly speaking, there is no such thing as a U.S. national housing market, unlike a centralized stock or futures market", yet finds – somewhat contradictory – that "there is a national factor that has contributed to an individual metropolitan area's housing price movement".

Clark and Coggin (2009) study convergence of US regional HPIs on Census Division level using OFHEO (FHFA) data with structured time series analysis methods. They identify two *super-* *regions*, as they call them, using principal component analysis. The first one includes East North Central, East South Central, Mountain and West North Central; and the second one: Midatlantic, New England, Pacific, South Atlantic and nation (also refer to Table 1 for the region codes used in this paper's tables and figures). Clark and Coggin (2009) found the evidence of the convergence in super-region 2 but not in super-region 1. Note, that CWSC was excluded as an outlier.

Holmes et al. (2011) study pair-wise convergence of US house prices on MSA level using Freddie Mac's Conventional Mortgage HPI data. They conduct pair-wise unit root tests on each MSA pair and collect the statistics on the rate of rejections. They find the extensive evidence of convergence of MSA HPIs, and also report the significance of the geographical distance between the regions.

Apergis and Payne (2012) use state-level FHFA HPI data employing a certain statistical clustering procedure, and find what they call *convergence clubs*. They report three such convergence clubs: the largest one with 29 states and the smallest one with only two states (Arkansas and Mississippi). In other words, there might be multiple equilibria and different regions may converge to one these equilibria.

Finally, I have to mention that there is an extensive literature on UK house price convergence. Meen (1999) brought to attention the notion of a *ripple effect*, i.e. the propensity for house prices to first rise in South East then spread across the rest of the country. He proposed the model where the regional house prices are in long-run relationship to each other. He tested the convergence of time series of ratios of house prices using the unit root test. Since this paper there were numerous studies of UK regional house prices using variety of techniques. While some authors find the evidence, others do not. I shall refer to Holmes and Grimes (2008) for the references to other research, and will only mention Cook (2003), who proposed to use a form of unit root testing of UK house price convergence that allows for asymmetric adjustments in the convergence process.

## Model

I study the convergence of regional house prices in the USA by testing for constancy of the ratio of the regional HPI to the national HPI. If the regional housing markets converge in the long-run then the time series of this ratio should be stationary, see Figure 1. Following Cook (2003) I shall use augmented momentum threshold autoregressive (MTAR) asymmetric unit root test to test for nonstationarity of the series of these ratios. I also produce the out-of-sample forecasts of regional house price appreciation series using the same model specification and compare them to the forecasts produced by simple ARMA(R,M) processes with and without an exogenous national house price appreciation inputs. If the regional house prices converge to the national index then including the national prices as exogenous variables should improve the forecasts produced by ARMA model which does not have exogenous national variables. Hence, comparison of out-of-sample forecasts produced by these models should provide additional support to the inferences drawn from in-sample convergence tests, such as unit root test.

### Unit root test

The basic idea of the unit root tests for convergence can be demonstrated with the standard form of the first order autoregressive process AR(1), see Shumway and Stoffer (2011) Section 3.2:

$$(1) \qquad x_t = \mu + \phi x_{t-1} + \varepsilon_t, \quad \varepsilon_t \sim \mathcal{N}(0, \sigma^2)$$

where $x_t$ – time series, $\mu$ – the intercept and $\varepsilon_t$ – error term. If $\phi = 1$ then this is not AR(1) process but a random walk process which is not stationary. Assuming $x_t = 0$ its mean and variance are given by the following equations, see Shumway and Stoffer (2011) Examples 1.14 and 1.18:

$$E[y_t] = \mu t$$

$$Var[x_t] = t\sigma^2$$

If $|\phi| < 1$ then I have AR(1) process that is stationary and its mean and variance are given by the following equations, see Shumway and Stoffer (2011) Section 3.2:

$$E[x_t] = \frac{\mu}{1 - \phi}$$

$$Var[x_t] = \frac{\sigma^2}{1 - \phi^2}$$

The case when $\phi = 1$ is referred to as "unit root", and it indicates a non-stationary process.

In order to form a hypothesis to test for unit root Equation 1 can be re-arranged as:

(2) $$\Delta x_t = \mu + \varphi x_{t-1} + \varepsilon_t$$

$$\Delta x_t = x_t - x_{t-1}$$

$$\varphi = \phi - 1$$

The null hypothesis is $(H_0: \varphi = 0)$, and the failure to reject it points to a unit root, see Dickey-Fuller (1979).

In order to accommodate higher order AR processes, the augmented DF, or ADF, test is usually performed. It has the same basic structure but it also contains lagged differences on the right hand side of the Equation (2). Since the house price series in our sample are quite persistent – as demonstrated by the autocorrelation function (ACF) plot on Figure 2 – ADF test should be better suited for the unit root testing, and as mentioned above it was used many times to study the convergence in UK regional house prices starting with Meen (1999) with inconclusive results with regards to the presence of convergence.

It was shown by Cook (2003) that the inconclusive results of application of ADF to UK housing market could be attributed to the fact that the adjustments to deviations from the constant ratio of regional to national HPI are asymmetrical. Therefore, in this work I shall use MTAR test, see Enders and Granger (1998), which accounts for such an asymmetry.

Following Cook (2003) I use the augmented form of MTAR test that is specified as follows:

(3) $$\Delta \tilde{y}_t = \beta_1 I_t \tilde{y}_{t-1} + \beta_2 (1 - I_t) \tilde{y}_{t-1} + \sum_{j=1}^{4} \beta_{2+j} \Delta \tilde{y}_{t-j} + \xi_t$$

$$\xi_t \sim \mathcal{N}(0, \sigma^2)$$

$$I_t = \begin{cases} 1, if\ \Delta \tilde{y}_{t-1} \geq 0 \\ 0, if\ \Delta \tilde{y}_{t-1} < 0 \end{cases}$$

$$y_t = ln\left(\frac{HPI_{reg}(t)}{HPI_{usa}(t)}\right)$$

$$\tilde{y}_t = y_t - \bar{y}$$

(4) $$\Delta \tilde{y}_t = \tilde{y}_t - \tilde{y}_{t-1}$$

Here, $y_t$ – a ratio of regional to national HPI, where both series are quarterly; $\bar{y}$ – estimate of the mean of series $y_t$; $\tilde{y}_t$ – "demeaned" series $y_t$; $reg$ – subscript that denotes a region studied, $I_t$ – a particular form of Heaviside indicator function; $\Delta y_t$ – the first difference series of the regional to national ratio, $\xi_t$ – error term. Asymmetric effects are captured by parameters $\beta_1, \beta_2$ as follows.

If the ratio of regional to national HPI increased in the previous period compared to the quarter before it then the Equation 3 will have a form: $\Delta \tilde{y}_t = \beta_1 \tilde{y}_{t-1} + \sum_{j=1}^{4} \beta_{2+j} \Delta \tilde{y}_{t-j} + \xi_t$ ; while if this ratio decreased then the Equation 3 will have a form: $\Delta \tilde{y}_t = \beta_2 \tilde{y}_{t-1} + \sum_{j=1}^{4} \beta_{2+j} \Delta \tilde{y}_{t-j} + \xi_t$. Compare this to Equation 2, where the single parameter $\varphi$ captures the response to both decreases and increases in the ratio of regional to national HPIs.

Note, that $\Delta \tilde{y}_t$ can also be re-written as:

(5) $$\Delta \tilde{y}_t = \Delta y_t = \Delta HPI_{reg}(t) - \Delta HPI_{usa}(t)$$

(6) $$HPA_i(t) \equiv \Delta HPI_i(t) = \ln\left(\frac{HPI_i(t)}{HPI_i(t-1)}\right)$$

where subscript $i$ corresponds to either a regional or a national series. In this form $\Delta y_t$ (and $\Delta \tilde{y}_t$) can be interpreted as a contemporaneous deviation of regional quarterly house price appreciation from national HPA (both defined by Equation 6).

The null hypothesis for unit root in this case is the simultaneous condition:

(7) $$H_0: \beta_1 = \beta_2 = 0$$

The test statistics is the usual F-statistics, but the critical values were computed by Monte-Carlo simulation and can be found in Enders (2001) Table 5, where the distribution of the statistics is referred to as $\Phi_\mu^*$. For the relevant range of 100 to 200 observations the critical values are given in Table 3, e.g. at 95% significance they are 4.72 to 4.71.

### Forecasting

Next, I use the same model specification given by Equations 3 for unit root testing to forecast the regional HPIs. At any given time $t$ all the terms on the right hand side of the main equation are known. For instance, since at the most recent observation $t = T$ the values $y_t, \Delta y_{t \leq T}$ are available, I can compute $I_{T+1}$ and the forecast $\Delta \hat{y}_{T+1}$. *Having the forecast of national HPI – consider the forward-looking national HPI scenario for stress testing –* I can compute $\Delta HPI_{usa}(t)$ for future periods, and obtain the regional HPA forecast using Equations 5:

(8) $$\Delta \widehat{HPI}_{reg}(t) = \Delta \hat{y}_t + \Delta HPI_{usa}(t), \quad t > T$$

I refer to $\Delta \widehat{HPI}_{reg}(t)$ as *MTAR-forecasts* in this paper.

I also model the regional HPA series in a standard ARMA(R,M) framework, see Shumway and Stoffer (2011) Section 3 as follows:

(9) $$\Delta HPI_{reg}(t) = \sum_{j=1}^{R} \phi_j \Delta HPI_{reg}(t-j) + \varepsilon_t + \sum_{j=1}^{M} \theta_j \varepsilon_{t-j}, \quad \varepsilon_t \sim \mathcal{N}(0, \sigma^2)$$

where $\Delta HPI_{reg}(t)$, given by Equation 6 – the quarterly house price appreciation for a region $reg$, $\phi_j$ – autoregressive coefficients, $\theta_j$ – moving average coefficients, and $\varepsilon_t$ – error terms.

The Equation 9 is my baseline model, in addition to which I specified the ARMAX(R,M) model that has a national quarterly HPA as an exogenous input in addition to ARMA parameters:

(10) $$\Delta HPI_{reg}(t) = \sum_{j=1}^{R} \phi_j \Delta HPI_{reg}(t-j) + \varepsilon_t + \sum_{j=1}^{M} \theta_j \varepsilon_{t-j} + \beta_1 \Delta HPI_{USA}(t)$$

$$\varepsilon_t \sim \mathcal{N}(0, \sigma^2)$$

where $\Delta HPI_{USA}(t)$, given by Equation 6 – the quarterly house price appreciation for the USA.

Summarizing, the baseline forecasts use only historical regional HPI time series, while MTAR and ARMAX forecasts have an exogenous national HPI series as inputs.

## Data and Empirical Results

### Data sources

There are several home price indices for US housing markets. For the national HPI I used the Excel file with historical series of macroeconomic variables from FRB CCAR web site, which also provides the projected HPIs to be used in stress testing. The National HPI variable is described in this data set as CoreLogic HPI seasonally adjusted by FRB staff.

I obtained the regional HPI time series from Moody's Analytics Economy.com service. The regional HPIs are maintained by Fiserv using Case-Shiller methodology, see S&P (2009), and are called "Fiserv Case-Shiller" HPI, see Fiserv CS HPI in References section. They can be located by mnemonic RHCSHP1TIQ in Economy.com Data Buffet service. The series were further seasonally adjusted by Moody's Analytcis using ARIMA-12 algorithm, see Moody's Analytcis (2011) and U.S. Census Bureau (2012). I downloaded data for 9 Census Divisions; the map of the regions is shown on Figure 3.

I used CoreLogic HPI for national and Case-Shiller HPI for regional series mainly because I did not have access to historical regional CoreLogic HPI series. This should not be a problem for the presented analysis since these indices "tend to move together because of their similar computation and included loan types" according to Noeth and Sengupta (2011), who compared widely used house indices. Moreover, I think that this would be a typical situation for many banks involved in FRB CCAR exercise because their internal models may have been estimated using indices other than CoreLogic, such as FHFA or Freddie Mac HPIs, subsequently they will have to somehow map the CoreLogic HPI-based scenarios to scenarios based on other indices.

Both sources of historical HPI data provide seasonally adjusted quarterly series from at least 1976 through 2012 Q2, see Table 2 with descriptive statistics of series. Our full sample contains 146 quarterly HPI observations from 1976 Q1 to 2012 Q2. The plots of the ratio of regional to national HPI series, i.e. $y_t$ as per Equations 3, are shown on Figure 1. All series converge in 2000 Q1 because at this period both CoreLogic and Case-Shiller HPI levels are normalized to 100 for all regions and the nation. This plot does not contradict the hypothesis of the convergence of regional HPI series: the ratios $y_t$ appear to be slowly meandering around 0.

### Unit root test

To test for the convergence of regional HPIs, first, I estimated the MTAR model parameters for each Census Division using a full sample. The dependent variable is a log difference of the ratio of regional to national HPI, as specified by Equation 4. All models came out with significant overall F-test statistics, but the results concerning the main subject of this paper vary. The model parameter estimates are shown in Table 3, as well as the corresponding test statistics of the MTAR unit root hypothesis test given by Equation 7 and their critical values.

In terms of unit root tests two Census Divisions at 90% confidence for four regions the unit root hypothesis can be rejected at 95% confidence with critical value between 3.69 and 3.81: Midatlantic, New England, Pacific and West North Central. These regions include 22 states and approximately 40% of US population as of 2012 according to US Census Buerau. I conclude that only four divisions show the evidence of regional HPI convergence to national HPI at 90% confidence while at 95% confidence New England drops out, and we are left with only three converging regions. Note, that Clark and Coggin (2009) and Apergis and Payne (2012) point to existence of clusters of convergence, possibly multiple equilibria, while Holmes et al. (2011) find the convergence to a single national HPI. My findings appear to support the latter for a significant part of the country, while no conclusions can be drawn for the rest of the USA in relation to the convergence or nonconvergence of regional HPI series.

Since the presence of serial autocorrelation in unit roots testing equation residuals casts doubt on its results, I produced autocorrelation and partial autocorrelation function plots at 95% confidence of the MTAR residuals for all 9 regions to ensure that the residuals do not have this problem, see Figure 4-Figure 12. Visual examination of plots does not show the evidence of the presence of significant serial correlation.

### Comparison of out-of-sample forecasts

This section describes the assessment of the forecasting power of the MTAR model and ARMAX model with exogenous national HPA input compared to a simple ARMA(R,M) model. The idea is to augment the findings of unit root testing with comparison of forecast produced with or without exogenous national HPI inputs. If the regional HPIs converge to national HPI then inclusion of exogenous national HPI inputs should improve out-of-sample forecasts compared to a simple ARMA model that uses only regional time series.

I produced the ex-post forecasts of regional HPA series using MTAR model as follows. The *training sample* was selected comprised of the observations from 1976 Q1 through 2008 Q4, and estimated MTAR model parameters. Then using these estimated parameters I produced the series of *dynamic* 1-, 4- and 8-step forecasts of the regional HPA variables $\Delta \widehat{HPI}_{reg}(t)$ according to Equation 8. The choice of 1-, 4- and 8-step forecasts was partially motivated by the fact that CCAR involved banks will have to produce one- and two year-ahead forecasts of their financials.

By *dynamic forecast* I mean the following: let us denote 2008 Q4 as $t = T$ and produce the h-step forecast, employing the procedure described for the Equation 8 and using the *actual* national

HPI series, but *not using* actual historical observations of *regional* HPI series beyond time $T$. I repeat this procedure setting $T$ equal to 2009 Q1, Q2, Q3 etc., until the end of the full sample is reached. For instance, for 1-step dynamic forecast I produce 14 predictions, 11 predictions of 4-step, and 7 predictions of 8-step dynamic forecast. To understand these settings consider the 8-step forecast starting at period $T = 2008$ Q4 that will produce the prediction for 2010 Q4. If I keep incrementing $T$ then after 6 more steps the prediction period will reach 2012 Q2 – our last available historical observation, therefore, with training sample ending at 2008 Q4 only seven 8-step forecasts can be produced that can be compared to actual numbers.

Next, I computed the root mean square forecast error (RMSFE) measure as follows:

$$RMSFE = \sqrt{\frac{1}{n}\sum_{i=1}^{n}(\hat{x}_i - x_i)^2}$$

where $\hat{x}_i, x_i$ – forecast and actual value of a variable. Note, that RMSFE is a relative measure for comparison between models, and it does not have a test statistics associated with it. Lower RMSFE means better forecasts. This measure can be seen as a combination of the accuracy (bias) and precision (variance) metrics. To see this consider the following re-arrangement of the above equation:

$$RMSFE^2 = (E[\hat{x}_i - x_i])^2 + Var[\hat{x}_i - x_i]$$

In this case our variable of interest is $\Delta\widehat{HPI}_{reg}(t)$, and the RMSFE results are reported in Table 4 for each h-step forecast set.

I compare these forecasts with ARMA(R,M) model's *h*-step dynamic forecasts, which were obtained in a similar way: produce *h*-step forecasts for each quarter starting from 2009 Q1, and again do not use any actual historical data beyond the time $t = T$. For regional ARMA(R,M) models I found the optimal values of autoregressive and moving average parameters R and M by comparing AIC statistics calculated from a set of trial (R,M) pairs, which were used to estimate the model parameters on a full sample, the model parameter estimates are shown in Table 5. Once optimal R and M were found they were used to estimate the model parameters on a training sample. Since I used the same starting point, i.e. 2009 Q1, for dynamic forecasts of regional HPA $\Delta \widehat{HPI}_{reg}(t)$, I had the same number of forecasts for the same periods as with the MTAR forecasts. The RMSFE performance of ARMA(R,M) model is shown in Table 4.

Similarly to ARMA I produced the ARMAX model (see Equation 10) forecasts, the only difference was that ARMAX model has an additional *exogenous* variable $\Delta HPI_{USA}(t)$. The model parameter estimates and the RMSFE results are shown in Table 6 and Table 4.

The columns on the right hand side of the Table 4 labeled "Winner" show the relative quality of MTAR, ARMAX and ARMA models in terms of RMSFE compared pair-wise, e.g. the third column from the right shows that ARMAX model produced better forecasts for all nine Census Divisions compared to ARMA. The second column from the right compares ARMA and MTAR models, and shows that for all regions except East South Central, New England and West South Central MTAR model had lower (better) RMSFE measure compared to ARMA. Out-of-sample forecast comparison shows that inclusion of national HPI produces better forecasts in both MTAR and ARMAX frameworks.

Finally, I performed the *forecast encompassing test* of ARMA and MTAR model forecasts, see Ericsson (1992) and Chong and Hendry (1986), and show the results in Table 7. The idea of the test is to regress the observed values on the predicted values from two models, then to test the linear restrictions of the following form:

(11) $$x_t = \alpha_0 + \alpha_1 \hat{x}_{1,t} + \alpha_2 \hat{x}_{2,t}$$

$$1: H_0: \{\alpha_0 = 0, \alpha_1 = 1, \alpha_2 = 0\}$$

$$2: H_0: \{\alpha_0 = 0, \alpha_1 = 0, \alpha_2 = 1\}$$

Here, $\hat{x}_{1,t}, \hat{x}_{2,t}$ – forecasts of regional HPA series $\Delta \widehat{HPI}_{reg}(t)$ obtained using MTAR and ARMA models, and $x_t$ – observed values of these series.

If the first null-hypothesis is not rejected while the second one is rejected, then the MTAR model forecast encompasses the ARMA forecast, i.e. ARMA forecast errors contain information that is at least partially explained by MTAR forecasts. Conversely, if the second null-hypothesis is not rejected while the first one is rejected, then ARMA forecast encompasses MTAR forecast. The results in Table 7 show that at 95% confidence MTAR forecast encompasses ARMA forecast for 5 Census Divisions, while ARMA encompasses MTAR forecast for only two regions: East and West South Central. Even for these two regions unlike the other regions the test results are inconclusive in two out of three *h*-step forecasts; hence, there is a stronger overall evidence of MTAR forecasts encompassing ARMA forecasts.

The regional HPA forecast plots are shown on Figure 13-Figure 21 for all 9 Census Divisions. Every plot has three rows: for 1-, 4- and 8-step dynamic forecasts. ARMA forecasts are plotted in the left column, and MTAR forecasts are in right column. The $2\sigma$ confidence bands (green) as well as the observed historical values (red) are shown too.

## Summary of findings and discussion

The results of unit root testing in Table 3 show the evidence of convergence at 90% confidence for four Census Divisions that include 22 states with over 40% of USA population: Midatlantic, New England, Pacific and West North Central. At 95% confidence the MTAR critical values are higher, so New England division fails to reject the unit root hypothesis.

The Census Division with the lowest F-statistics in unit root hypothesis testing is South Atlantic (see Table 3). It includes West Virginia, Maryland, Delaware, Virginia, North Carolina, South Carolina, Georgia, Florida states and the District of Columbia. Based on a separate analysis of Virginia state, not presented in this paper, I saw that its HPI was on a substantially different path than a national HPI and states like Florida in recent years. I speculate that this may have been caused by the disproportional allocation of government spending in this region. For instance, Sauter et al. (2012) list Virginia (#2), Marylans (#3) and West Virginia (#9) in the top ten states receiving most federal money through defense procurement, Medicare and other large item spending channels in 2010. Further analysis is necessary to pinpoint the reason why South Atlantic division is not showing the long-run convergence to national HPI, which may require the convergence testing on state or MSA level.

Comparing $\beta_1$ and $\beta_2$ parameter estimates in Table 3 for the regions which show the convergence one can see that for Midatlantic and West North Central divisions one of the parameters is more significant than the other in terms of t-statistics and the parameter values are quite different. This observation points to a possible asymmetry in the response to the deviation of the regional growth rate from the national HPA. I plan to test the asymmetry hypothesis in further study. Remember that the unit root hypothesis defined by Equation 7 only tests for the joint significance of $\beta_1$ and $\beta_2$, and asymmetry requires a separate test.

In out-of-sample forecast comparisons based on RMSFE measures MTAR model performed better than ARMA for all regions except New England, East and West South Central divisions, while ARMAX model was an overall winner. The ARMAX model appears to be better suited for forecasting regional HPIs in short-term (up to years) than both ARMA and MTAR models. Since MTAR specification given by Equation 3 was constructed to test long-run relationship between regional and national HPI series, it is probably failing to capture some short-run features of regional series. Perhaps addition of moving average terms, like in ARMA framework, would improve MTAR model's forecasting power.

In forecast encompassing tests MTAR model encompassed ARMA for East North Central, Mountain, Pacific, South Atlantic and West North Central divisions; while ARMA encompassed MTAR forecasts only for East and West South Central division, and only in one out of three *h*-step forecasts.

In practical settings 4- and 8-step forecasts are most important for they correspond to one or two-year-ahead forecasts, which are required by regulator routinely. However, note that some items in financial reporting such as loan loss provisions and impairments may require life-of-loan cash flow projections, i.e. extremely long term forecasts of HPI series. In the 8-step forecasts MTAR performed worse in terms of RMSFE only in three regions: West South/North Central, East South Central divisions.

Out-of-sample dynamic forecasting uses model parameter estimates on a training sample including observations through 2008 Q4, then predicts the quarterly regional house price appreciation rates for the forecasting period starting on 2009 Q1. The *dynamic* feature of the forecast reflects the fact that these forecasts do not use actual values in the forecasts. For instance while generating 8-step forecast of regional HPA in 2012 Q2 the last historical *regional* HPI observation used was 2010 Q2, while the model parameter estimates were still from 2008 Q2. Thus the dynamic *h*-step out-of-sample forecasts emulate actual use of the forecasting models in practice, and my analysis should reflect properly the forecasting power of the models considered.

Overall the tests results are consistent with each other. The convergence of regional house prices implies that national HPI should have information about regional HPIs, hence the MTAR and ARMAX models are expected to perform better than ARMA, and, generally, they do so. One anomaly is that for the regions, where unit root hypothesis was not rejected, MTAR still has performed better in out-of-sample forecasts compared to ARMA.

All the analysis was performed using MATLAB 2012b and its econometrics, financial time series and statistical toolboxes. The MTAR model was estimated using `LinearModel` function. The ARMA models were estimated with `garchfit` function. The dynamic h-step forecasts were produced by our own custom scripts.

## Conclusion

I found the evidence of regional US house price convergence across the 40% (in terms of population) of the country: the asymmetric unit root test of the ratio of regional to national HPI series points to existence of convergence in four Census Divisions out of nine. Comparison with ARMA forecasts supports this finding. This comparison is relevant because the simple ARMA model does not use national HPI data. Hence, if national HPI contains the information about regional prices then MTAR (and ARMAX) forecasts, which use national HPI as an exogenous input, must be more accurate.

MTAR model forecasts had lower RMSFE, i.e. better combination of accuracy and precision, than ARMA model for 2/3 of the regions. The forecast encompassing tests show that MTAR forecasts encompass ARMA forecast for five regions, which is consistent with RMSFE comparison: the regions, which had higher RMSFE, are either encompassed by ARMA or yield inconclusive results in encompassing tests. ARMAX model forecasts were better overall in terms of RMSFE, thus it appears to be better suited for the purpose of the short-term (up to two years-ahead) forecasting.

As I mentioned in introduction one of the primary motivations for this study was the inclusion of the national HPI among 26 other macroeconomic variables in the stress testing scenarios of FRB CCAR. These scenarios provide three sets of projections, which are intended for use within the annual stress testing exercise: baseline, stress and adverse stress scenario. Each scenario contains quarterly projection of all variables through 2015. FRB does not provide any guidance on disaggregation of the national HPI into regional series.

Taking into an account that the regional house prices do not appear to converge in the long-run to national HPI in almost 60% of the country I conclude that although providing the national HPI projections in stress scenarios is not completely meaningless – the national HPI appears to contain information about regional house price moves in some parts of the USA – the national to regional disaggregation of HPI is not a trivial exercise. I believe that banks take different approaches to this problem, and FRB may end up collecting the data, which is based on very different and possibly inconsistent assumptions from participating banks.

Taking into account the relative size of the mortgage books in bank's assets, this clearly is an issue for both the banks implementing the stress scenarios and the regulator interpreting the results. It will be even more serious next year when smaller banks, i.e. 10-50 billion dollars in total assets, will be required to conduct the stress testing. While the nation's largest banks may have resources to develop proprietary regional HPI models, their smaller peers may not. At the same time, smaller banks are also more likely to have geographically concentrated mortgage loan books making it even more important for them to have sensible regional HPI models.

## List of references

# Tables

| Code | Census Division | Census Region |
|---|---|---|
| CENC | East North Central Division | Midwest |
| CESC | East South Central Division | South |
| CMAC | Middle Atlantic Division | Northeast |
| CMTN | Mountain Division | West |
| CNEC | New England Division | Northeast |
| CPAC | Pacific Division | West |
| CSAC | South Atlantic Division | South |
| CWSC | West South Central Division | South |
| CWNC | West North Central Division | Midwest |

Table 1 Census Divisions and the codes used in this paper

| | USA | | CENC | CESC | CMAC | CMTN | CNEC | CPAC | CSAC | CWSC | CWNC |
|---|---|---|---|---|---|---|---|---|---|---|---|
| | HPI | | | | | | | | | | |
| Count | 146 | | 146 | 146 | 146 | 146 | 146 | 146 | 146 | 146 | 146 |
| Min | 21.81 | | 25.46 | 36.81 | 21.88 | 28.16 | 16.67 | 19.32 | 36.27 | 36.82 | 30.65 |
| Max | 201.61 | | 143.09 | 135.2 | 212 | 194.35 | 185.04 | 237.25 | 214.09 | 143.8 | 162.72 |
| Last | 142.26 | | 102.65 | 119.92 | 166.11 | 113.21 | 149.88 | 136.4 | 134.46 | 143.8 | 119.84 |
| | Quarterly HPA (see Equation 6) | | | | | | | | | | |
| Count | 145 | | 145 | 145 | 145 | 145 | 145 | 145 | 145 | 145 | 145 |
| Mean | 1.29% | | 0.96% | 0.81% | 1.40% | 0.96% | 1.51% | 1.35% | 0.90% | 0.94% | 0.94% |
| Median | 1.44% | | 1.21% | 0.90% | 1.16% | 0.94% | 1.19% | 1.66% | 1.03% | 0.91% | 1.10% |
| Min | -5.44% | | -5.67% | -4.30% | -3.14% | -11.07% | -3.31% | -10.87% | -7.46% | -3.63% | -6.31% |
| Max | 5.93% | | 4.60% | 7.03% | 7.52% | 7.32% | 9.04% | 6.71% | 5.97% | 5.40% | 4.81% |
| Last | 2.04% | | 2.63% | 0.28% | 1.10% | 6.53% | 1.24% | 3.69% | 1.57% | 1.44% | 1.51% |
| Std.dev. | 1.97% | | 1.65% | 1.31% | 2.12% | 2.78% | 2.27% | 3.09% | 2.07% | 1.35% | 1.80% |
| Skew | -0.76 | | -1.27 | 0.35 | 0.27 | -1.19 | 0.50 | -1.44 | -1.18 | 0.39 | -1.19 |
| Kurt | 1.68 | | 2.75 | 5.69 | -0.25 | 4.49 | 0.60 | 3.63 | 3.34 | 2.27 | 2.56 |

Table 2 Descriptive statistics of HPI series and their growth rates for USA and 9 Census Divisions

| Parameters (see Eq.3) | CENC | CESC | CMAC | CMTN | CNEC | CPAC | CSAC | CWSC | CWNC |
|---|---|---|---|---|---|---|---|---|---|
| $\beta_1$ | -0.0026 | -0.0047* | -0.0464*** | -0.0043 | -0.0266*** | -0.0586*** | -0.0119* | -0.0016 | 0.0044 |
| $\beta_2$ | 0.1205* | 0.1192* | 0.1092* | 0.3949*** | 0.3649*** | 0.4165*** | 0.1282** | 0.486*** | 0.4362*** |
| $\beta_3$ | 0.352*** | 0.1915** | 0.3222*** | 0.1194* | 0.2738*** | 0.3736*** | 0.2101*** | 0.157** | 0.0119 |
| $\beta_4$ | 0.2981*** | 0.3156*** | 0.2582*** | 0.2356*** | 0.2736*** | 0.0453 | 0.1531** | 0.3141*** | 0.062* |
| $\beta_5$ | -0.072* | 0.1207* | 0.0317 | -0.0894* | -0.1074* | -0.141** | 0.055* | -0.167** | 0.0327 |
| $\beta_6$ | -0.0176** | 0.0006 | -0.0267** | -0.01* | -0.0205* | -0.0105* | -0.0153** | -0.0569*** | -0.01* |
| R^2 | 0.33 | 0.42 | 0.37 | 0.34 | 0.54 | 0.48 | 0.22 | 0.60 | 0.36 |
| Adj.R^2 | 0.30 | 0.40 | 0.35 | 0.32 | 0.52 | 0.46 | 0.19 | 0.59 | 0.33 |
| F-stat (see Eq.7) | 2.37 | 1.96 | 6.49 | 2.85 | 4.46 | 4.93 | 1.25 | 2.82 | 9.06 |

| Crit. val. | Conf. | # of obs. | | Crit. val. | Conf. | # of obs. | |
|---|---|---|---|---|---|---|---|
| 3.81 | 90% | 100 | | 4.72 | 95% | 100 | |
| 3.69 | 90% | 200 | | 4.71 | 95% | 200 | |

Table 3 MTAR parameter estimates and unit root test statistics (*,**,*** - 1-,2- and 3-$\sigma$ significance) and critical values from Table 5 in Enders (2001). Unit root rejected at 90% (grey cells) and at 95% (bold blue figures) confidence.

| | 1-step | | | 4-step | | | 8-step | | | Winner | | |
|---|---|---|---|---|---|---|---|---|---|---|---|---|
| Region | ARMA | ARMAX | MTAR | ARMA | ARMAX | MTAR | ARMA | ARMAX | MTAR | ARMA vs ARMAX | ARMA vs MTAR | ARMAX vs MTAR |
| CENC | 0.0213 | 0.0163 | 0.0147 | 0.0232 | 0.0158 | 0.0187 | 0.0215 | 0.0187 | 0.0114 | ARMAX | MTAR | ARMAX |
| CESC | 0.0176 | 0.0107 | 0.0147 | 0.0146 | 0.0101 | 0.0195 | 0.0105 | 0.0195 | 0.0195 | ARMAX | ARMA | ARMAX |
| CMAC | 0.0153 | 0.0100 | 0.0146 | 0.0218 | 0.0106 | 0.0187 | 0.0296 | 0.0187 | 0.0149 | ARMAX | MTAR | ARMAX |
| CMTN | 0.0419 | 0.0298 | 0.0249 | 0.0566 | 0.0326 | 0.0250 | 0.0506 | 0.0250 | 0.0287 | ARMAX | MTAR | MTAR |
| CNEC | 0.0105 | 0.0076 | 0.0122 | 0.0132 | 0.0091 | 0.0156 | 0.0141 | 0.0156 | 0.0127 | ARMAX | ARMA | ARMAX |
| CPAC | 0.0257 | 0.0170 | 0.0183 | 0.0466 | 0.0166 | 0.0265 | 0.0355 | 0.0265 | 0.0152 | ARMAX | MTAR | ARMAX |
| CSAC | 0.0245 | 0.0135 | 0.0091 | 0.0293 | 0.0101 | 0.0085 | 0.0250 | 0.0085 | 0.0072 | ARMAX | MTAR | MTAR |
| CWSC | 0.0088 | 0.0078 | 0.0218 | 0.0110 | 0.0104 | 0.0253 | 0.0103 | 0.0253 | 0.0199 | ARMAX | ARMA | ARMAX |
| CWNC | 0.0192 | 0.0163 | 0.0148 | 0.0345 | 0.0256 | 0.0151 | 0.0114 | 0.0151 | 0.0182 | ARMAX | MTAR | MTAR |

Table 4 Root mean squared forecast error (RMSFE) for three models

| Parameters | CENC | CESC | CMAC | CMTN | CNEC | CPAC | CSAC | CWSC | CWNC |
|---|---|---|---|---|---|---|---|---|---|
| Intercept | 0.0028** | 0.0016* | 0.0004** | 0.0148*** | 0.0041* | 0.0053*** | 0.0003*** | 0.0054** | 0.0039** |
| $\phi_1$ | -0.9463*** | 0.3115* | 1.8035*** | -0.4265** | 0.0842* | 1.0766*** | 1.2049*** | -0.4711*** | 0.883*** |
| $\phi_2$ | 0.8342*** | 0.4843*** | -0.8335*** | -0.2568*** | -0.0224 | -1.0179*** | 0.4845* | -0.1656*** | -0.945*** |
| $\phi_3$ | 0.7936*** | | | 0.1331* | 0.6669*** | 0.8564*** | -0.712*** | 0.5186*** | 0.6336*** |
| $\phi_4$ | | | | 0.6593*** | | -0.1544*** | | 0.7297*** | |
| $\theta_1$ | 1.4533*** | -0.2123* | -1.5782*** | 1.117*** | 0.4803*** | -0.0854*** | -0.6452*** | 0.9867*** | 0.0371** |
| $\theta_2$ | -0.0934* | -0.2705** | 0.7168*** | 1.1574*** | 0.5086*** | 0.9988*** | -0.8442*** | 0.7763*** | 1*** |
| $\theta_3$ | -0.5467*** | 0.3928*** | | 0.9211*** | | | 0.4948*** | 0.2843** | |
| $\theta_4$ | | -0.2974*** | | | | | | -0.4263*** | |
| $\sigma^2$ | 0.0001*** | 0.0001*** | 0.0002*** | 0.0003*** | 0.0002*** | 0.0002*** | 0.0002*** | 0.0001*** | 0.0001*** |

Table 5 ARMA model parameter estimates (*,**,*** - 1-,2- and 3-$\sigma$ significance), see Eq.(9)

| Parameters | CENC | CESC | CMAC | CMTN | CNEC | CPAC | CSAC | CWSC | CWNC |
|---|---|---|---|---|---|---|---|---|---|
| Intercept | 0.0007* | 0.0032** | 0.0013 | -0.0039** | 0.0025* | -0.0021* | -0.0029** | 0.0018* | 0.0005 |
| $\phi_1$ | 0.5338*** |  | 0.3962*** | 0.2973*** | 0.5844*** | 0.7575*** | 0.2921*** | -0.497*** | 0.2525*** |
| $\phi_2$ | -0.5784*** |  | -0.3954*** | -0.3625*** |  | -0.3821*** | -0.275*** | 0.422*** | 0.0547* |
| $\phi_3$ | 0.5362*** |  | 0.6346*** | 0.3003*** |  | 0.1951** | 0.2008*** | 0.4347*** | -0.1445* |
| $\phi_4$ | -0.2369** |  |  |  |  | -0.1333** |  | 0.2869*** | 0.2142** |
| $\theta_1$ | -0.4766*** | -0.0828* | -0.2226** | 0.1009* | -0.1115* | -0.19** | -0.0605* | 1*** | 0.5254*** |
| $\theta_2$ | 0.8133*** | 0.1111** | 0.7677*** | 0.5964*** | 0.127** | 0.4204*** | 0.4003*** |  | 0.3192** |
| $\theta_3$ | -0.5386*** | 0.2997*** | -0.4011*** | 0.1089* | 0.3749*** |  |  |  | 0.0787 |
| $\theta_4$ | 0.4698*** |  |  | -0.1767** |  |  |  |  | -0.4086*** |
| $\beta_1$ | 0.0001*** | 0.3756*** | 0.3003*** | 0.8923*** | 0.3029*** | 0.7605*** | 0.7686*** | 0.1157*** | 0.4152*** |
| $\sigma^2$ | 0.4928*** | 0.0001*** | 0.0002*** | 0.0002*** | 0.0002*** | 0.0001*** | 0.0001*** | 0.0001*** | 0.0001*** |

Table 6 ARMAX model parameter estimates (*,**,*** - 1-,2- and 3-σ significance), see Eq.(10)

| Region | Encompassing model | | | |
|---|---|---|---|---|
|  | 1-step | 4-step | 8-step | Overall |
| CENC | MTAR | MTAR | MTAR | MTAR |
| CESC | Inconclusive | Inconclusive | ARMA | ARMA |
| CMAC | Inconclusive | Inconclusive | Inconclusive | Inconclusive |
| CMTN | MTAR | MTAR | Inconclusive | MTAR |
| CNEC | Inconclusive | Inconclusive | Inconclusive | Inconclusive |
| CPAC | MTAR | MTAR | MTAR | MTAR |
| CSAC | MTAR | MTAR | MTAR | MTAR |
| CWSC | ARMA | Inconclusive | Inconclusive | ARMA |
| CWNC | MTAR | MTAR | Inconclusive | MTAR |

Table 7 Forecast encompassing tests between ARMA and MTAR model forecasts at 95% confidence (see Eq.9) for *h*-step dynamic forecasts
25

# Figures

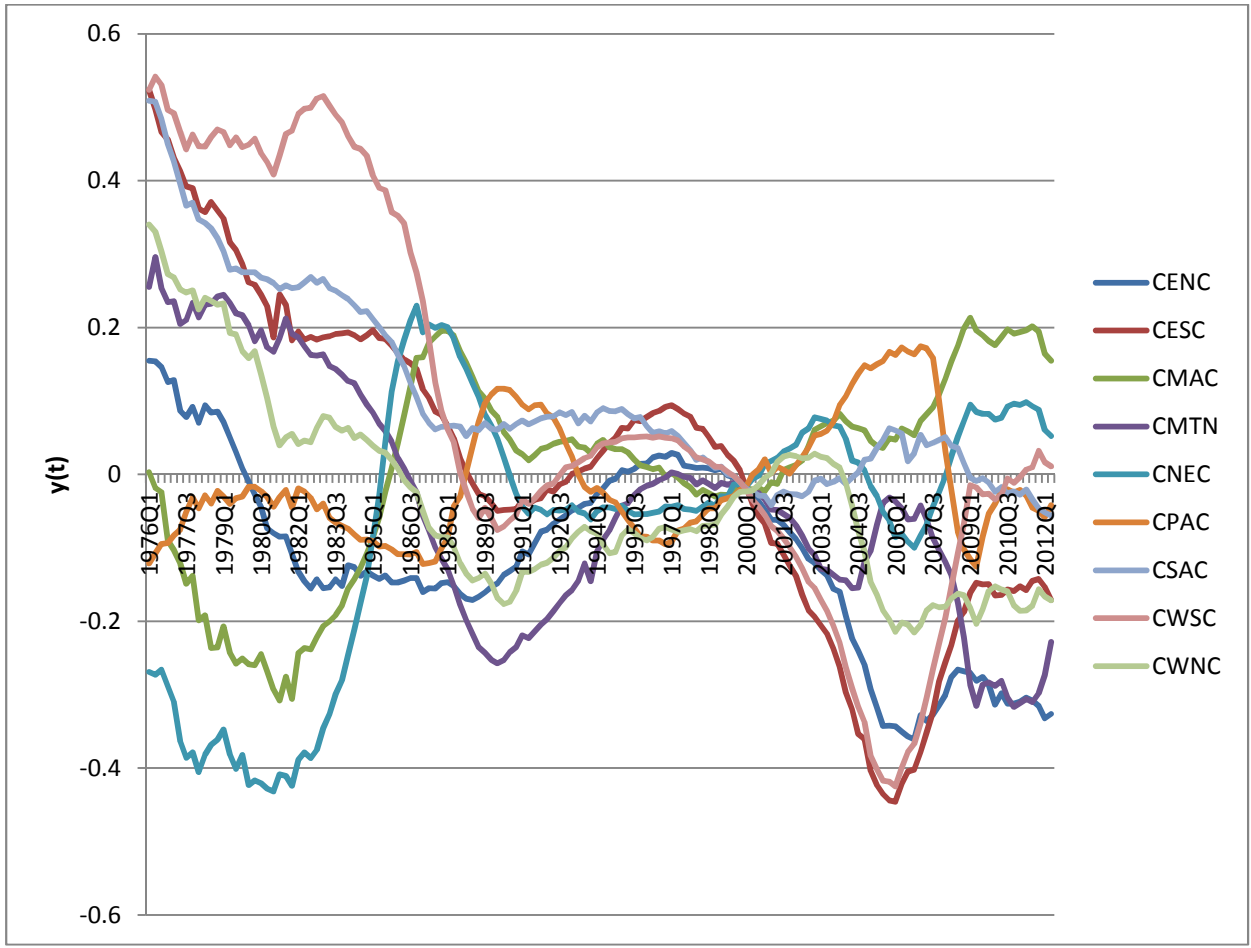

Figure 1 Logarithm of ratio of regional HPI to national HPI for 9 Census Divisions (see Eq.3)

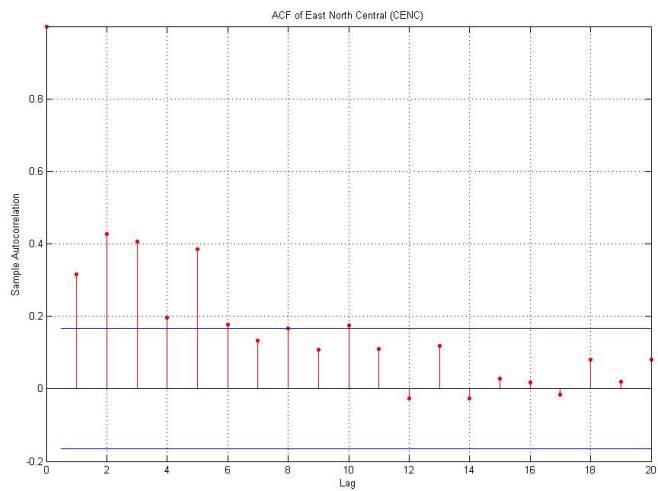

Figure 2 Autocorrelation plot of $\Delta y_t$ (see Eq.4) for East North Central Census Division

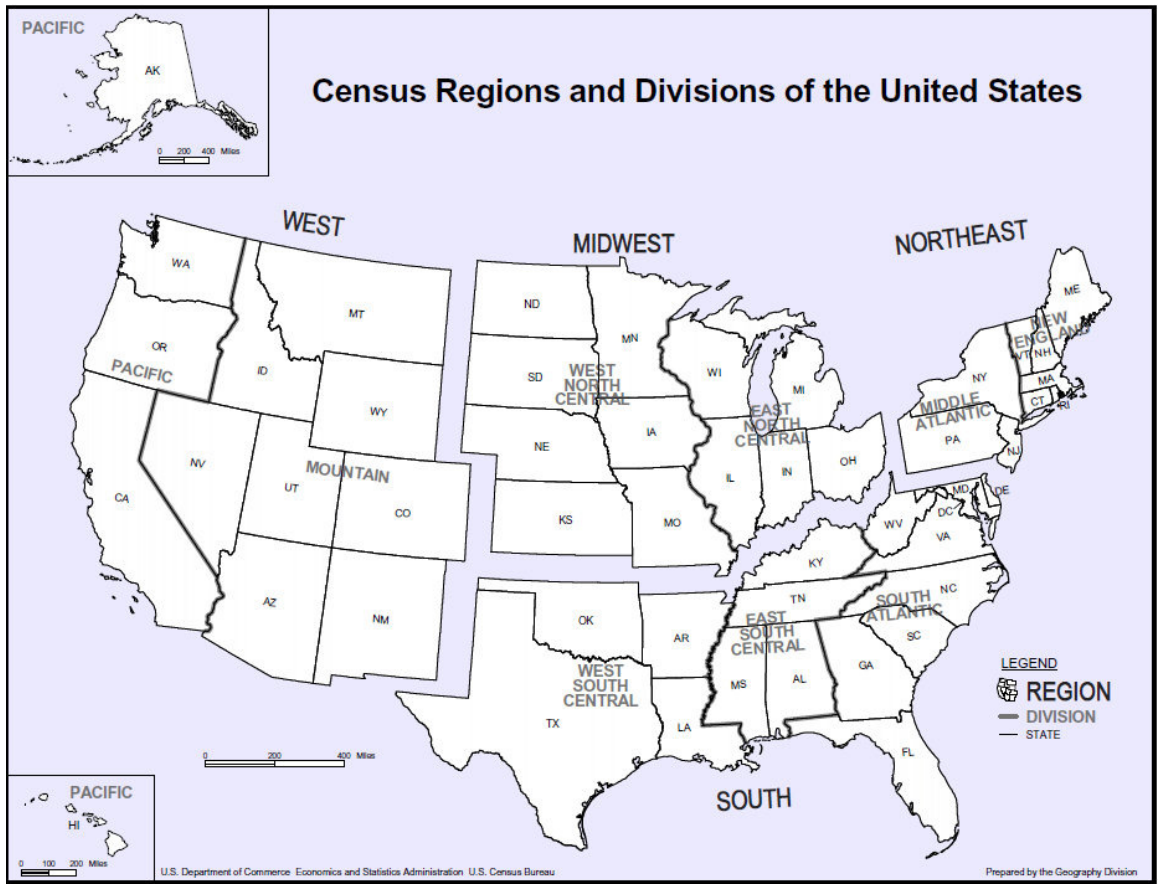

**Figure 3 Map of US Census Divisions**

## ACF and PACF of MTAR residuals

The following set of 9 figures shows autocorrelation (ACF) and partial autocorrelation (PACF) plots of residuals $\xi_t$ from MTAR Equation 3.

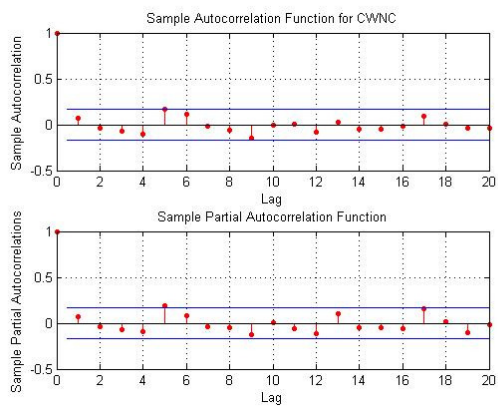

**Figure 4 ACF and PACF of MTAR residuals for West North Central Census Division**

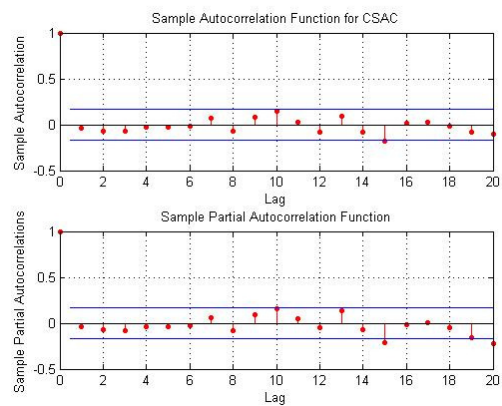

**Figure 5 ACF and PACF of MTAR residuals for South Atlantic Census Division**

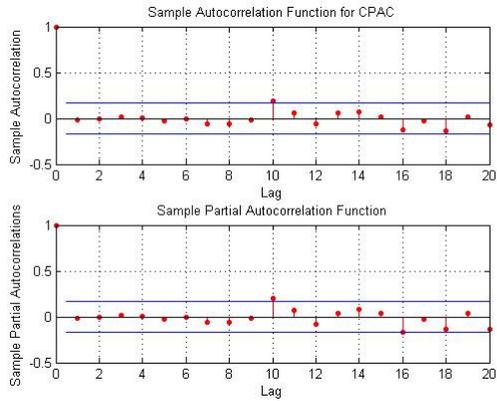

**Figure 6** ACF and PACF of MTAR residuals for Pacific Census Division

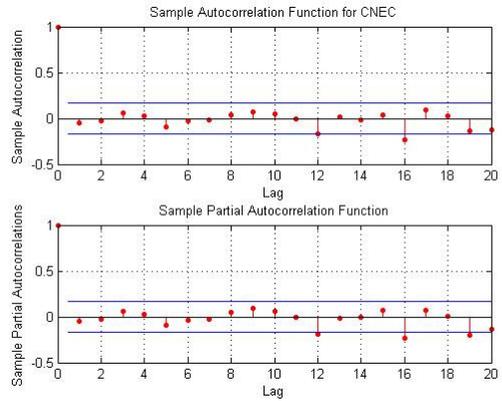

**Figure 7** ACF and PACF of MTAR residuals for New England Census Division

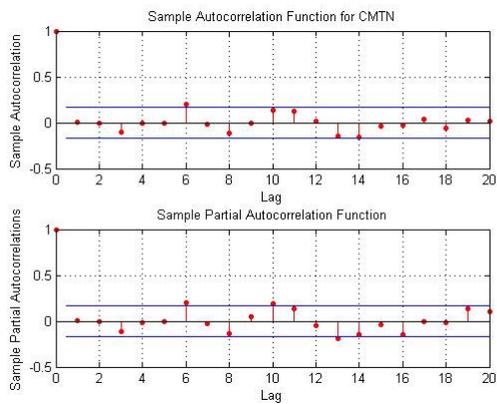

**Figure 8** ACF and PACF of MTAR residuals for Mountain Census Division

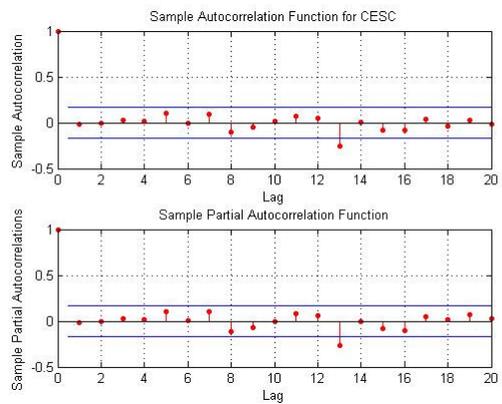

**Figure 9** ACF and PACF of MTAR residuals for East South Central Census Division

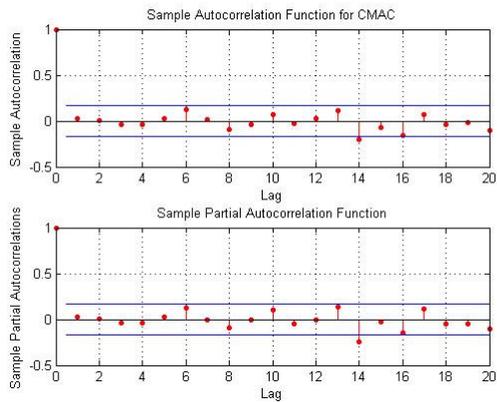

**Figure 10** ACF and PACF of MTAR residuals for Midatlantic Census Division

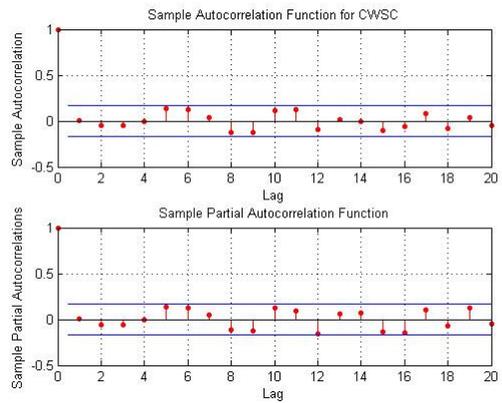

**Figure 11** ACF and PACF of MTAR residuals for West South Central Census Division

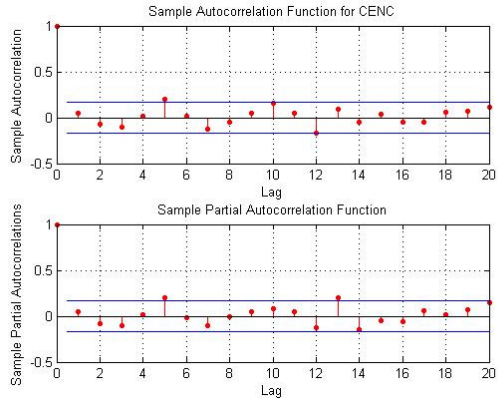

**Figure 12 ACF and PACF of MTAR residuals for East North Central Census Division**

## Out-of-sample forecasts: ARMA vs. MTAR

The following set of figures is out-of-sample HPA forecast plots (blue lines) of ARMA (left) and MTAR (right) models for nine Census Divisions. The vertical line shows the end of the training sample. Red line is actual historical observations of HPA (Eq.6). Green lines are $2\sigma$ confidence bands. Three rows are for 1-, 4- and 8-step forecasts.

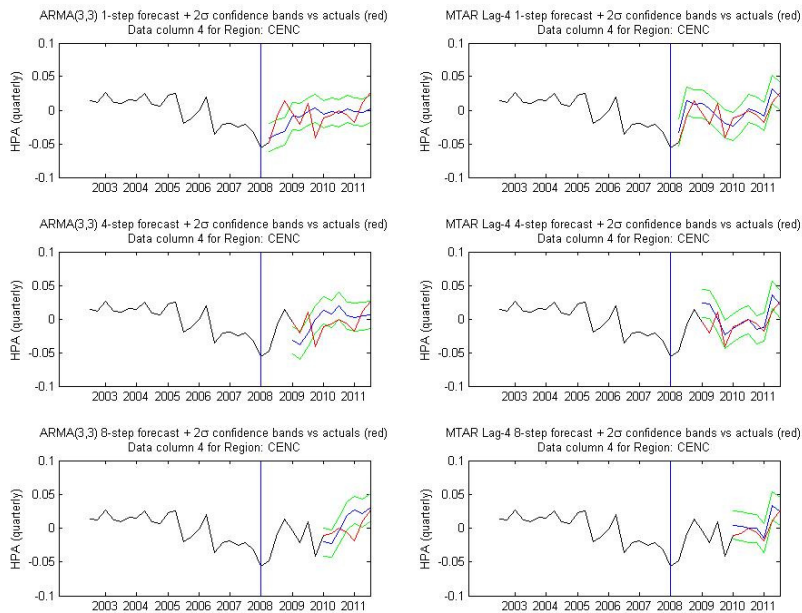

**Figure 13 ARMA and MTAR dynamic 1-, 4- and 8-step forecasts of quarterly house price appreciation series for East North Central Census Division**

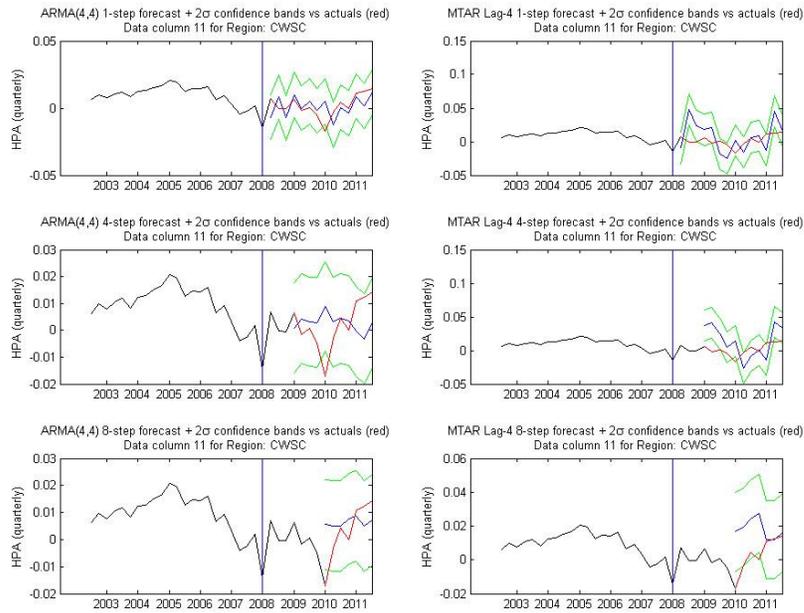

**Figure 14 ARMA and MTAR dynamic 1-, 4- and 8-step forecasts of quarterly house price appreciation series for West South Central Census Division**

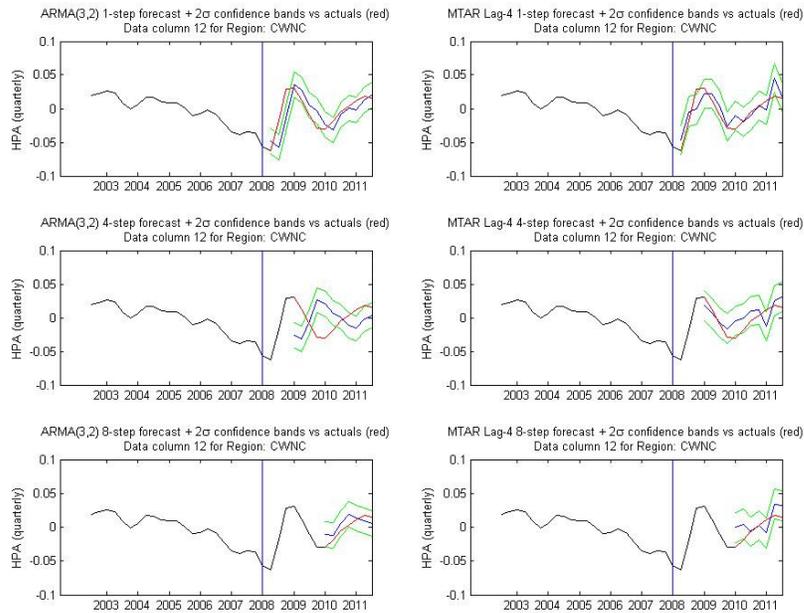

**Figure 15 ARMA and MTAR dynamic 1-, 4- and 8-step forecasts of quarterly house price appreciation series for West North Central Census Division**

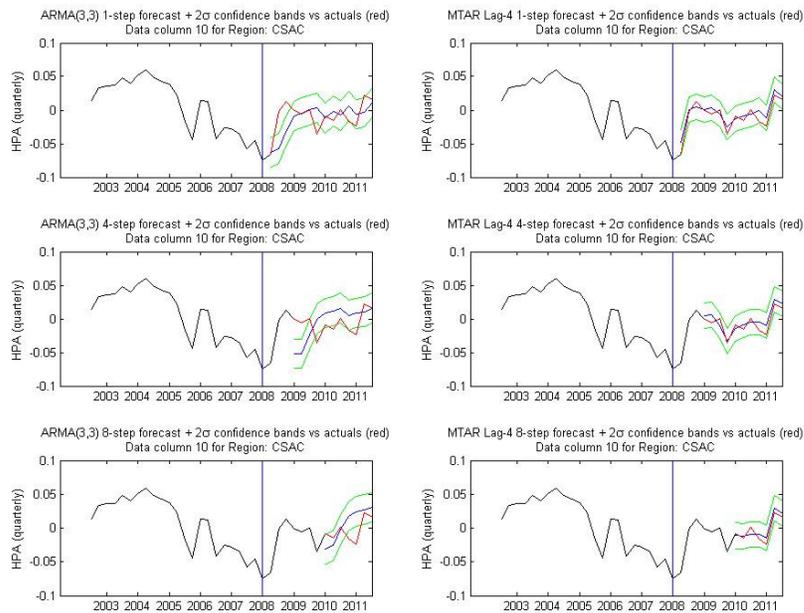

**Figure 16 ARMA and MTAR dynamic 1-, 4- and 8-step forecasts of quarterly house price appreciation series for South Atlantic Census Division**

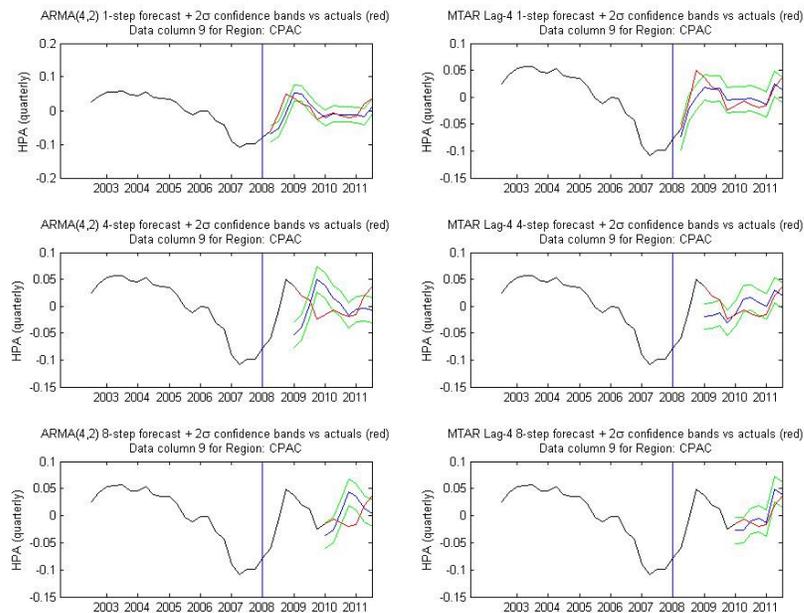

**Figure 17 ARMA and MTAR dynamic 1-, 4- and 8-step forecasts of quarterly house price appreciation series for Pacific Census Division**

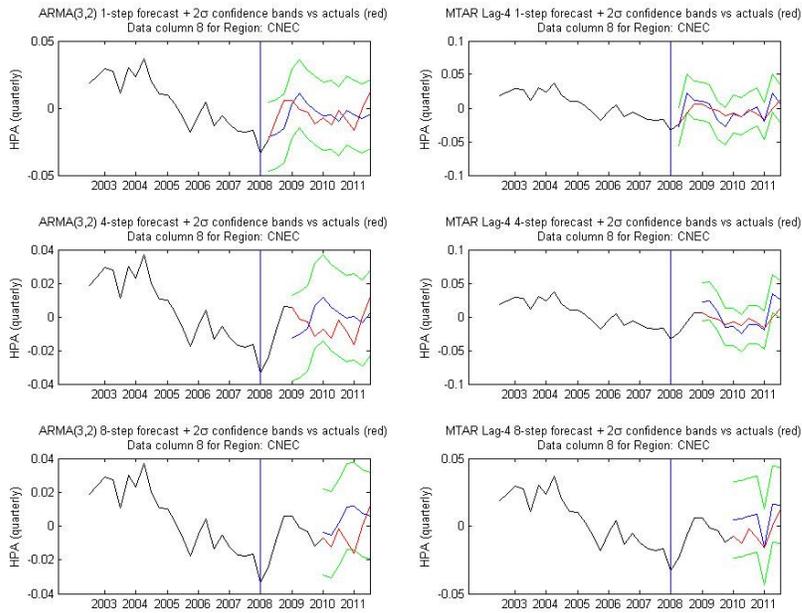

**Figure 18 ARMA and MTAR dynamic 1-, 4- and 8-step forecasts of quarterly house price appreciation series for New England Census Division**

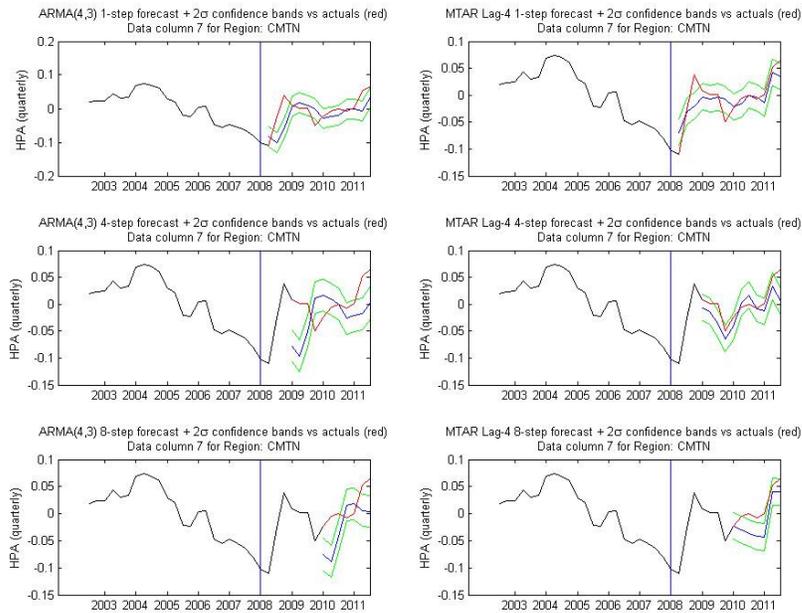

**Figure 19 ARMA and MTAR dynamic 1-, 4- and 8-step forecasts of quarterly house price appreciation series for Mountain Census Division**

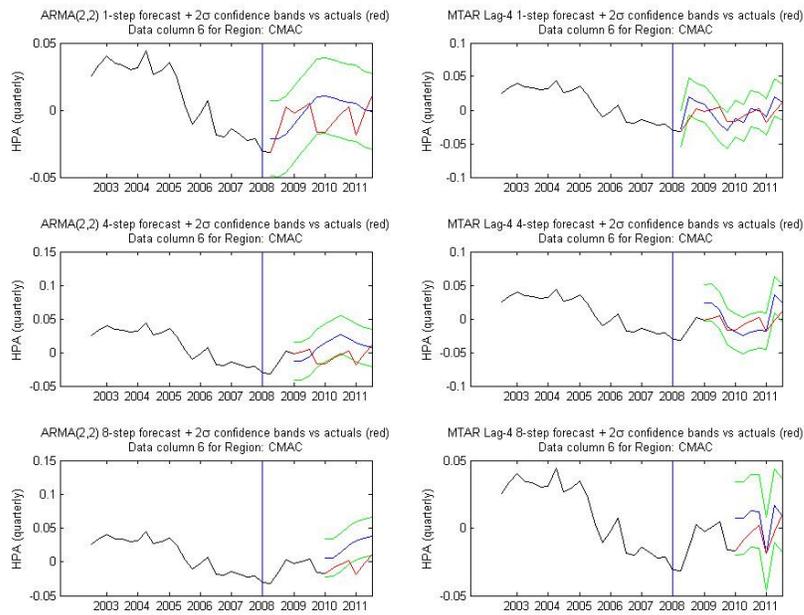

**Figure 20 ARMA and MTAR dynamic 1-, 4- and 8-step forecasts of quarterly house price appreciation series for Midatlantic Census Division**

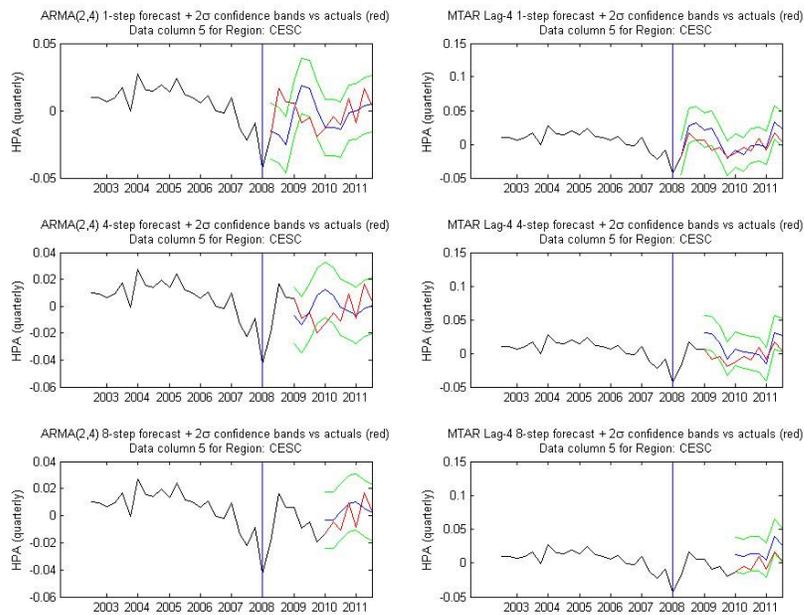

**Figure 21 ARMA and MTAR dynamic 1-, 4- and 8-step forecasts of quarterly house price appreciation series for East South Central Census Division**

## Out-of-sample forecasts: ARMA vs. ARMAX

The following set of figures is out-of-sample HPA forecast plots (blue lines) of ARMA (left) and ARMAX (right) models for nine Census Divisions. The vertical line shows the end of the training sample. Red line is actual historical observations of HPA (Eq.6). Green lines are $2\sigma$ confidence bands. Three rows are for 1-, 4- and 8-step forecasts.

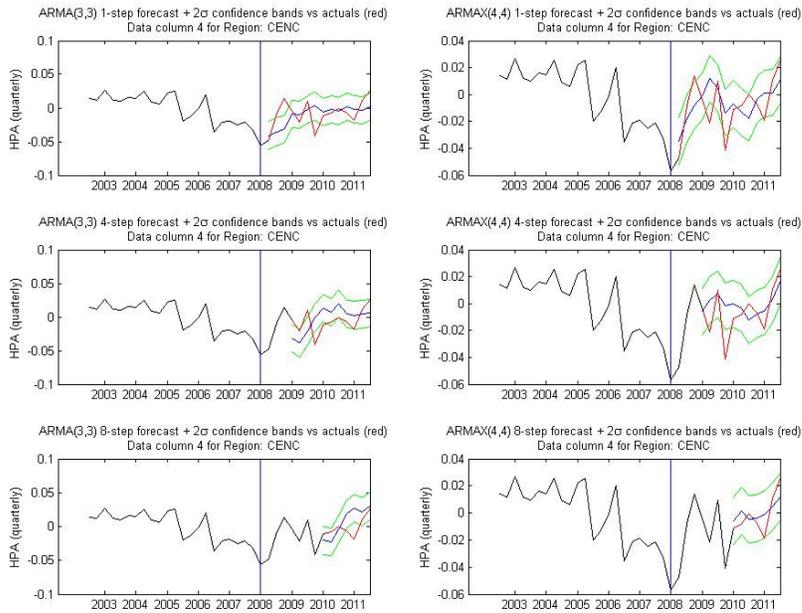

**Figure 22 ARMA and ARMAX dynamic 1-, 4- and 8-step forecasts of quarterly house price appreciation series for East North Central Census Division**

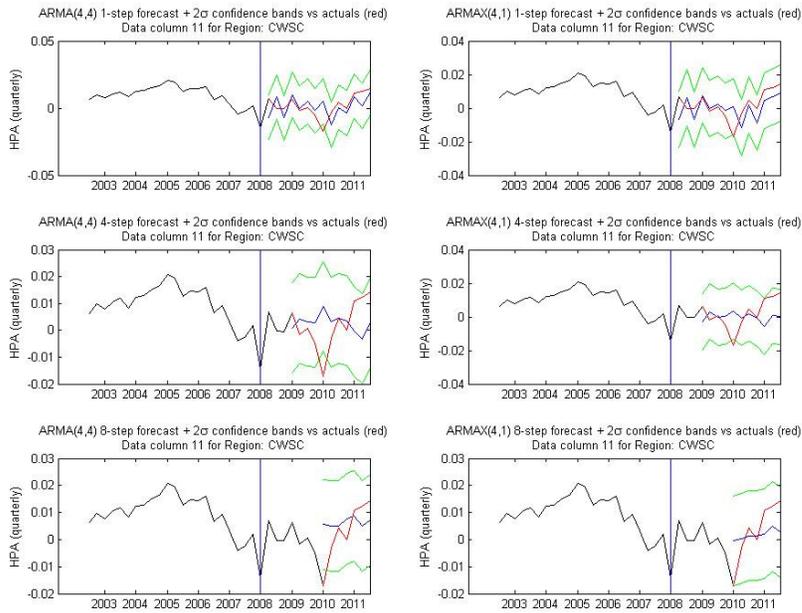

**Figure 23 ARMA and ARMAX dynamic 1-, 4- and 8-step forecasts of quarterly house price appreciation series for West South Central Census Division**

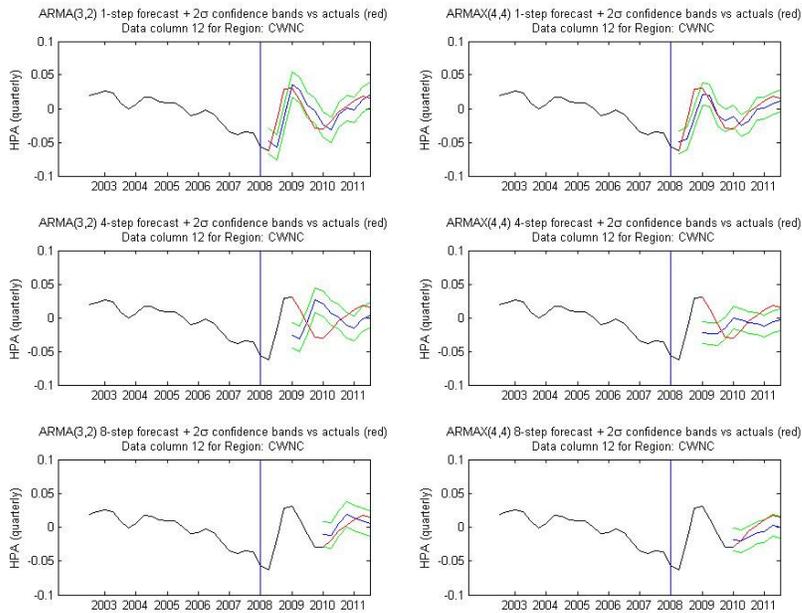

**Figure 24 ARMA and ARMAX dynamic 1-, 4- and 8-step forecasts of quarterly house price appreciation series for West North Central Census Division**

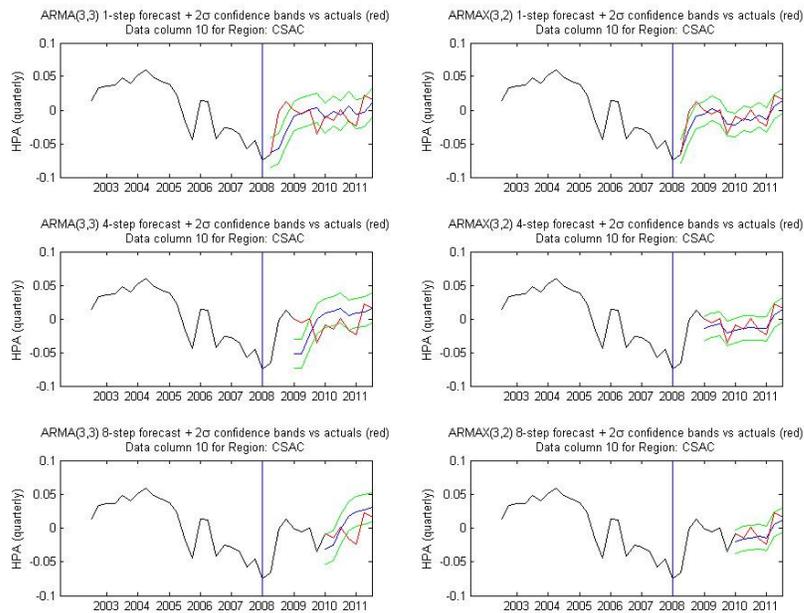

**Figure 25 ARMA and ARMAX dynamic 1-, 4- and 8-step forecasts of quarterly house price appreciation series for South Atlantic Census Division**

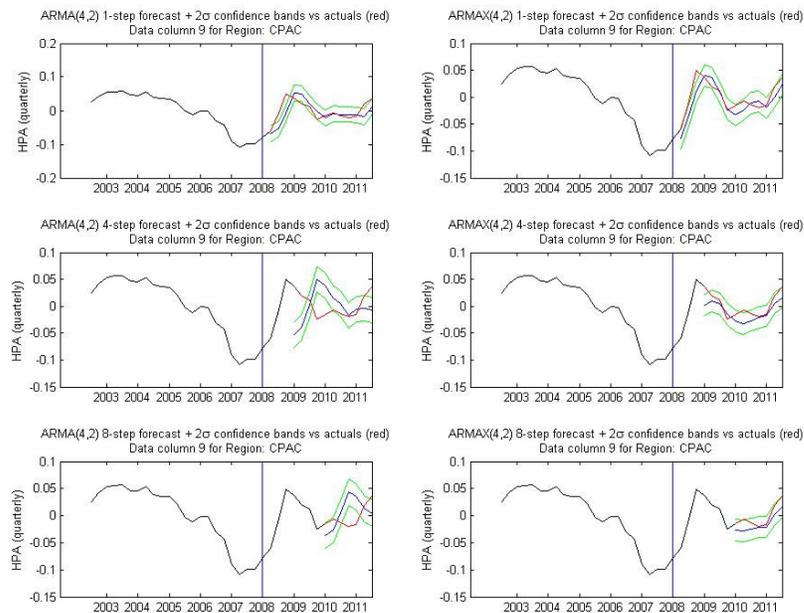

**Figure 26 ARMA and ARMAX dynamic 1-, 4- and 8-step forecasts of quarterly house price appreciation series for Pacific Census Division**

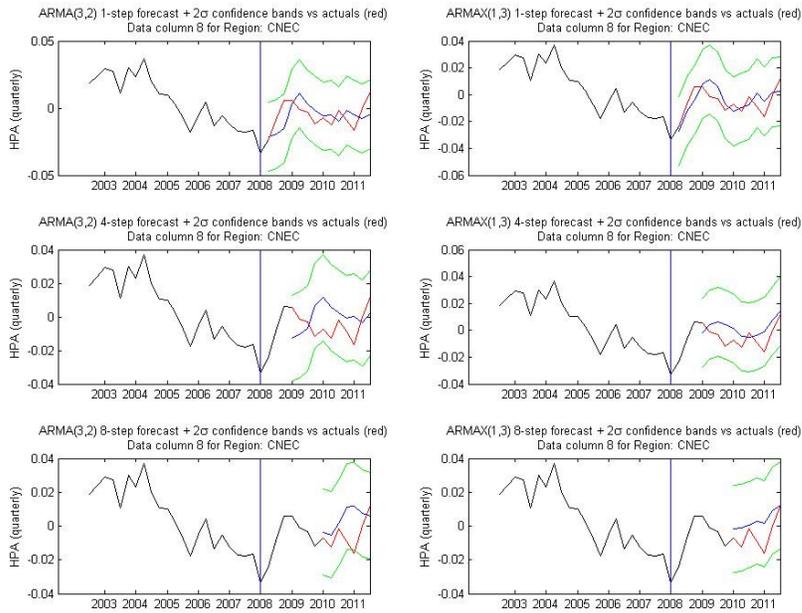

**Figure 27 ARMA and ARMAX dynamic 1-, 4- and 8-step forecasts of quarterly house price appreciation series for New England Census Division**

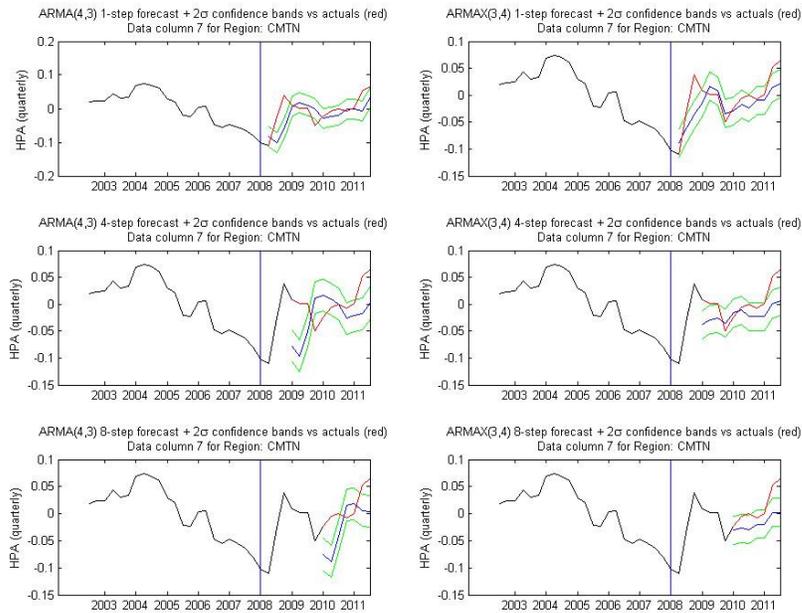

**Figure 28 ARMA and ARMAX dynamic 1-, 4- and 8-step forecasts of quarterly house price appreciation series for Mountain Census Division**

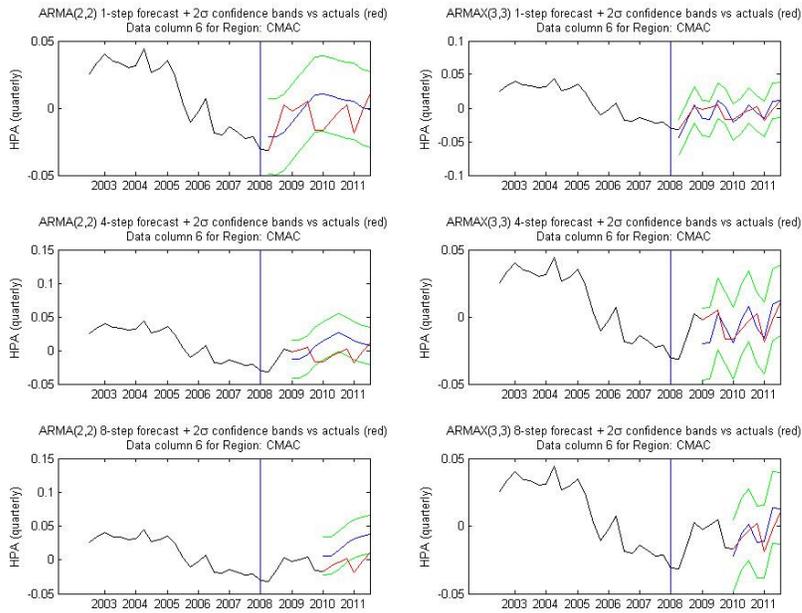

**Figure 29 ARMA and ARMAX dynamic 1-, 4- and 8-step forecasts of quarterly house price appreciation series for Midatlantic Census Division**

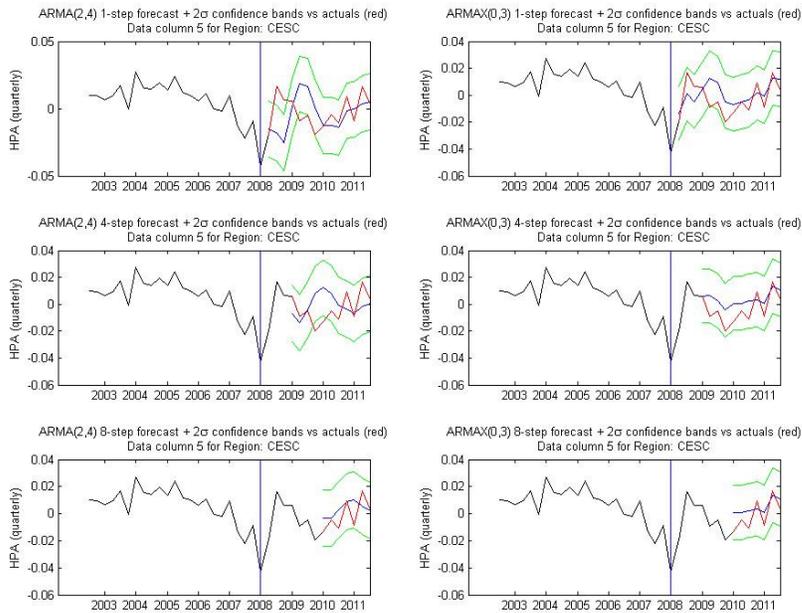

**Figure 30 ARMA and MTAR dynamic 1-, 4- and 8-step forecasts of quarterly house price appreciation series for East South Central Census Division**